\documentstyle[aps,12pt]{revtex}

\begin{document}
\title{A reductionistic approach to quantum computation}
\author{Giuseppe Castagnoli\thanks{%
Information Technology Dept., Elsag Bailey, 16154 Genova, Italy}, Dalida
Monti\thanks{%
Universit\`{a} di Genova and Elsag Bailey, 16154 Genova, Italy}}
\maketitle

\begin{abstract}
In the reductionistic approach, mechanisms are divided into simpler parts
interconnected in some standard way (e.g. by a mechanical transmission). We
explore the possibility of porting reductionism in quantum operations.
Conceptually, first parts are made independent of each other by assuming
that all ``transmissions'' are removed. The overall state would thus become
a superposition of tensor products of the eigenstates of the independent
parts. Transmissions are restored by projecting off all the tensor products
which violate them. This would be performed by particle statistics; the
plausibility of this scheme is based on the interpretation of particle
statistics as projection. The problem of the satisfiability of a Boolean
network is approached in this way. This form of quantum reductionism appears
to be able of taming the quantum whole without clipping its richness.
\end{abstract}

\date{}

\section{Introduction}

Reducing the apparently more complex to the less so is the aim of the
reductionistic approach. In (classical) applied mechanics, this can be
exemplified by the notion of {\em mechanical} {\em transmission}. At the
same time, transmissions divide the whole into simpler parts and reconstruct
it $-$ they introduce a ``divide and conquer'' strategy. In fig. 1, a
crank-shaft is the mechanical transmission which imposes a correlation
between the positions of parts $r$ and $s$ (here discretized as 0 and 1).

\medskip \medskip \medskip

\begin{center}
Fig. 1

\medskip
\end{center}

Things can be more difficult in quantum mechanisms, when the mechanical
transmission can become an interaction Hamiltonian which does not commute
with the Hamiltonians of the simpler parts. One can make this comparison:

\begin{itemize}
\item  Classical mechanisms:

\begin{itemize}
\item  advantage: the whole is approachable by reductionism, there is a
long-standing tradition of taming it through a divide and conquer strategy;

\item  disadvantage: the ``richness'' of the overall functionality grows
(roughly) polynomially with the number of parts.
\end{itemize}

\item  Quantum mechanisms:

\begin{itemize}
\item  disadvantage: the whole is, in general, difficult to tame without
clipping its richness;

\item  advantage: the ``richness'' of the overall functionality can grow
exponentially with the number of parts; this is the case of the quantum
algorithms found so far.$^{\left[ 1,2,3,4,5,6,7\right] }$
\end{itemize}
\end{itemize}

Is there a way of getting the advantages and avoiding the disadvantages of
either kind of mechanism? Namely: is the quantum whole tamable through a
divide and conquer strategy?

The answer given in this work is affirmative, although speculative. By
applying a sort of reverse engineering, we will identify a feature that
would be nice to have in quantum mechanisms. Using a top down approach might
be justified since the problem at stake, beyond its physical implications,
is an engineering one. The nice-to-have thing will be a ``quantum mechanical
transmission'' which, at the same time, divides the whole into simpler parts
and reconstructs it; it will be inspired by a special interpretation of
identical particles (parts) statistics. The form of reductionism thus
introduced turns out to be deeper than the classical one.

Then we ask ourselves whether or not the feature is physical. The answer
reached in this work is that it is physical as long as Hermitean matrices
can stand in place of Hamiltonians.

\section{Definition of quantum mechanical transmission}

We consider the mechanism of fig. 1. The ``quantum transmission'' should
establish a {\em constraint} between two quantum parts which would otherwise
be independent: part $r$ with eigenstates $\left| 0\right\rangle _{r}$ and $%
\left| 1\right\rangle _{r}$, and part $s$ with eigenstates $\left|
0\right\rangle _{s}$ and $\left| 1\right\rangle _{s}$ (fig. 2). We claim
that the allowed states of this quantum transmission have the form

\begin{equation}
\left| \varphi \right\rangle =\alpha \left| 0\right\rangle _{r}\left|
1\right\rangle _{s}+\beta \left| 1\right\rangle _{r}\left| 0\right\rangle
_{s}\text{ with }\left| \alpha \right| ^{2}+\left| \beta \right| ^{2}=1.
\end{equation}
We should note that the eigenvalues of {\em each} tensor product in (1)
satisfy the invertible Boolean NOT function.

\medskip

\medskip

\medskip

\begin{center}
Fig. 2
\end{center}

Let us assume that the transmission is temporarily removed. The generic
state of each independent part is

\[
\left| \Psi \right\rangle _{r}=\alpha _{r}\left| 0\right\rangle _{r}+\beta
_{r}\left| 1\right\rangle _{r},\ \left| \Psi \right\rangle _{s}=\alpha
_{s}\left| 0\right\rangle _{s}+\beta _{s}\left| 1\right\rangle _{s} 
\]

The whole, unentangled, state in the Hilbert space

\[
H=span\left\{ \left| 0\right\rangle _{r}\left| 0\right\rangle _{s},\left|
0\right\rangle _{r}\left| 1\right\rangle _{s},\left| 1\right\rangle
_{r}\left| 0\right\rangle _{s},\left| 1\right\rangle _{r}\left|
1\right\rangle _{s}\right\} ,\text{ is} 
\]

\begin{equation}
\left| \Psi \right\rangle =\alpha _{0}\left| 0\right\rangle _{r}\left|
0\right\rangle _{s}+\alpha _{1}\left| 0\right\rangle _{r}\left|
1\right\rangle _{s}+\alpha _{2}\left| 1\right\rangle _{r}\left|
0\right\rangle _{s}+\alpha _{3}\left| 1\right\rangle _{r}\left|
1\right\rangle _{s},
\end{equation}
with $\alpha _{o}=\alpha _{r}\alpha _{s}$, etc. Some tensor products violate
the NOT\ function (here also called ``symmetry'') characterizing this
quantum transmission, some do not. The transmission is restored by
projecting the compound state $\left| \Psi \right\rangle $ on the
``symmetric'' subspace $H_{s}=span\left\{ \left| 0\right\rangle _{r}\left|
1\right\rangle _{s},\left| 1\right\rangle _{r}\left| 0\right\rangle
_{s}\right\} .$ Let us define the ``symmetrization projection'' $A_{rs\text{ 
}}$by the following equations:

\[
A_{rs}\left| 0\right\rangle _{r}\left| 1\right\rangle _{s}=\left|
0\right\rangle _{r}\left| 1\right\rangle _{s},\ A_{rs}\left| 1\right\rangle
_{r}\left| 0\right\rangle _{s}=\left| 1\right\rangle _{r}\left|
0\right\rangle _{s},\text{ }A_{rs}\left| 0\right\rangle _{r}\left|
0\right\rangle _{s}=0,\ A_{rs}\left| 1\right\rangle _{r}\left|
1\right\rangle _{s}=0 
\]

Given $\left| \Psi \right\rangle $ (eq. 2), its projection on $H_{s}$ is
obtained (in a somewhat peculiar way which will be clearly motivated in
Section IV) by submitting a {\em free normalized vector} $\left| \varphi
\right\rangle $ of $H$ (whose amplitudes on the basis vectors of $H$ are
free and independent variables up to normalization) to the following,
mathematically simultaneous, conditions:

\begin{enumerate}
\item[i)]  $A_{rs}\left| \varphi \right\rangle =\left| \varphi \right\rangle 
$

\item[ii)]  the distance between the vector before projection $\left| \Psi
\right\rangle $ and that after projection $\left| \varphi \right\rangle $%
should be minimum.
\end{enumerate}

\noindent This yields the usual result $\left| \varphi \right\rangle
=k\left( \alpha _{1}\left| 0\right\rangle _{r}\left| 1\right\rangle
_{s}+\alpha _{2}\left| 1\right\rangle _{r}\left| 0\right\rangle _{s}\right) $%
, where $k$ is the renormalization factor, which is of course an allowed
transmission state.

A continuous operation on the state of either {\em part} $r$ or $s$,
associated with simultaneous, continuous projection of the {\em whole} state
on $H_{s}$ [conditions (i) and (ii)] will perform the quantum mechanical
transmission. This is inspired by a special interpretation of particle
statistics. See also ref. $\left[ 8,9\right] $.

\section{Particle statistics as a projection operator}

Let

\begin{equation}
\left| \Psi \left( t\right) \right\rangle =\alpha \left| 0\right\rangle
_{1}\left| 0\right\rangle _{2}+\beta \left| 1\right\rangle _{1}\left|
1\right\rangle _{2}+\gamma \left( \left| 0\right\rangle _{1}\left|
1\right\rangle _{2}+\left| 1\right\rangle _{1}\left| 0\right\rangle
_{2}\right)
\end{equation}
be a triplet state at some time $t$, where 1 and 2 are two identical
spinless particles; $0/1$ stand for, say, horizontal/vertical polarization.
There is a well known didascalic way of deriving the symmetrical form of
state (3). Particles 1 and 2 are first considered to be independent:

\[
\left| \Psi \right\rangle _{1}^{^{\prime }}=\alpha _{1}\left| 0\right\rangle
_{1}+\beta _{1}\left| 1\right\rangle _{1},\ \left| \Psi \right\rangle
_{2}^{^{\prime }}=\alpha _{2}\left| 0\right\rangle _{2}+\beta _{2}\left|
1\right\rangle _{2}. 
\]

Their compound, unentangled, state is

\[
\left| \Psi \right\rangle ^{^{\prime }}=\gamma _{0}\left| 0\right\rangle
_{1}\left| 0\right\rangle _{2}+\gamma _{1}\left| 0\right\rangle _{1}\left|
1\right\rangle _{2}+\gamma _{2}\left| 1\right\rangle _{1}\left|
0\right\rangle _{2}+\gamma _{3}\left| 1\right\rangle _{1}\left|
1\right\rangle _{2}, 
\]
where $\gamma _{0}=\alpha _{1}\alpha _{2},$ etc. This state must be
submitted to {\em symmetrization} for the permutation of the two particles,
namely to the operator $S_{12}=1+P_{12}$ followed by renormalization, and
this indeed yields a (entangled) triplet state like that of eq. (3).

The interpretation of particle statistics as the result of projection
amounts to taking this didascalic procedure literally. Symmetry is viewed as
the result of a {\em watchdog} {\em effect} which continuously {\em projects}
the overall state of the two particles on the three-dimensional Hilbert space

\[
H_{t}=span\left\{ \left| 0\right\rangle _{1}\left| 0\right\rangle
_{2},\left| 1\right\rangle _{1}\left| 1\right\rangle _{2},\frac{1}{\sqrt{2}}
\left( \left| 0\right\rangle _{1}\left| 1\right\rangle _{2}+\left|
1\right\rangle _{1}\left| 0\right\rangle _{2}\right) \right\} . 
\]
Therefore $\left| \Psi \left( t\right) \right\rangle $ symmetry is the
result of the continuous projection

\begin{equation}
\forall t:S_{12}\left| \Psi \left( t\right) \right\rangle =\left| \Psi
\left( t\right) \right\rangle .
\end{equation}

This is different from saying that a particle statistics symmetry is an {\em %
initial} {\em condition}, that was there from the beginning and has been
conserved as a constant of motion. When symmetry is viewed as the result of
projection, no independent initial condition is required. Any initial state
would itself be the result of a projection satisfying eq. (4), that is
satisfying the symmetry. As for the rest, this projection interpretation
appears to be consistent with the conventional interpretation. However, the
notion of a quantum mechanical transmission is a peculiar case, and will
have to rely strictly on the projection interpretation.

By way of exemplification, we shall illustrate a sort of quantum mechanical
transmission simply related to particle statistics. Let 1 and 2 be two free,
identical and non-interacting spin 1/2 particles. At a given time, the
overall spatial wave function $\Psi \left( x_{1,}x_{2}\right) $ is a linear
combination of the spatial wave functions of the two free particles: $\Psi
\left( x_{1}\right) =e^{ik_{A}x_{1}},\Psi \left( x_{2}\right)
=e^{ik_{B}x_{2}};$ $x_{1\text{ }}$and $x_{2}$ are the spatial coordinates of
the two particles:

\[
\Psi \left( x_{1,}x_{2}\right) =e^{ik_{A}x_{1}}e^{ik_{B}x_{2}}\pm
e^{ik_{A}x_{2}}e^{ik_{B}x_{1}}, 
\]
the + ($-$) sign goes with the spin singlet (triplet) state (normalization
is disregarded). It can be seen that

\medskip

$\left\| \Psi \left( x_{1,}x_{2}\right) \right\| ^{2}=\cos ^{2}kx,\ $for the
singlet state,

\medskip

$\left\| \Psi \left( x_{1,}x_{2}\right) \right\| ^{2}=\sin ^{2}kx,\ $for the
triplet state,

\medskip

where $x=x_{1}-x_{2},\ k=k_{A}-k_{B}$ (see fig. 3).

\medskip

\medskip

\medskip

\begin{center}
Fig. 3
\end{center}

For example, in the surrounding of the origin, close (separated) particles
are more likely to be found in a singlet (triplet) state. There is a sort of
quantum mechanical transmission inducing a correlation between the mutual
distance of the two particles and the character of their spin state. This is
of course a consequence of the entanglement between spatial and spin wave
functions created by particle statistics.

``Divide and conquer'' would work as follows. At the same time the whole is
divided into parts (the non-interacting particles) and reconstructed by
particle statistics, or projection. Noticeably, this quantum mechanical
transmission would fall apart if the two particles were not identical.

\section{Behaviour of the quantum mechanical transmission}

Let us go back to the transmission $r$,$s$ defined in Section II. We
consider an operation performed on just one qubit, say, $r.$ Let $\left|
0\right\rangle _{i}\equiv \left( 
\begin{tabular}{r}
$1$ \\ 
$0$%
\end{tabular}
\right) _{i}$, $\left| 1\right\rangle _{i}=\left( 
\begin{tabular}{r}
$0$ \\ 
$1$%
\end{tabular}
\right) _{i}$ with $i=r,s$, $Q_{r}\left( \varphi \right) \equiv \left( 
\begin{tabular}{rr}
$\cos \varphi $ & $-\sin \varphi $ \\ 
$\sin \varphi $ & $\cos \varphi $%
\end{tabular}
\right) _{r}$. We should think $Q_{r}\left( \varphi \right) $ to be a
continuous rotation, with $\varphi =\omega t$ and time $t$ ranging from 0 to 
$\frac{\varphi }{\omega }$.

Let us examine the effect of applying $Q_{r}\left( \varphi \right) $ to
qubit $r$,

\begin{equation}
\rho _{r}\left( t\right) =Q_{r}\left( \omega t\right) \rho _{r}\left(
0\right) Q_{r}^{\dagger }\left( \omega t\right) ,
\end{equation}
{\em under continuous} $A_{rs}$ {\em projection} of the overall state; $\rho
_{r}\left( t\right) $ is qubit $r$ density matrix at time $t$, we will be
concerned with the interval $0\leq \varphi \leq \frac{\pi }{2}$.

Let the initial state of the transmission be the ``symmetrical'' state
(whose tensor products satisfy the Boolean NOT, or symmetry $A_{rs}$):

\begin{equation}
\left| \Psi \left( 0\right) \right\rangle =\cos \vartheta \left|
0\right\rangle _{r}\left| 1\right\rangle _{s}+\sin \vartheta \left|
1\right\rangle _{r}\left| 0\right\rangle _{s}.
\end{equation}
For all times $t$, the state of the transmission is obtained by submitting a 
{\em free normalized vector} $\left| \Psi \left( t\right) \right\rangle $ of
the Hilbert space $H$ (Secion II) to the following {\em mathematically
simultaneous} conditions (by a rotation of $\rho _{r}$ under {\em continuous}
$A_{rs}$ projection, we assume that the two operations are {\em %
mathematically} {\em simultaneous }$-$ see ref. 8,9{\em \ }); therefore, 
{\em for all t }or $\varphi $:

\begin{enumerate}
\item[i)]  $A_{rs}\left| \Psi \left( t\right) \right\rangle =\left| \Psi
\left( t\right) \right\rangle ,$

\item[ii)]  $\rho _{r}\left( t\right) =Tr_{s}\left( \left| \Psi \left(
t\right) \right\rangle \left\langle \Psi \left( t\right) \right| \right)
=\cos ^{2}\left( \vartheta +\varphi \right) \left| 0\right\rangle
_{r}\left\langle 0\right| _{r}+\sin ^{2}\left( \vartheta +\varphi \right)
\left| 1\right\rangle _{r}\left\langle 1\right| _{r}$, where $Tr_{s}$ means
partial trace over $s,$

\item[iii)]  the distance between the vectors before and after projection
must be minimum. This is a condition moving in time together with the
continuous projection; the notion of the minimization of such a distance is
clearer by thinking in terms of finite increments.
\end{enumerate}

Conditions (i) and (ii) yield

\[
\left| \Psi \left( t\right) \right\rangle =\cos \left( \vartheta +\varphi
\right) \left| 0\right\rangle _{r}\left| 1\right\rangle _{s}+e^{i\delta
}\sin \left( \vartheta +\varphi \right) \left| 1\right\rangle _{r}\left|
0\right\rangle _{s}, 
\]
where $\delta $ is a still unconstrained phase, as can be readily checked;
condition (iii), given the transmission initial state (6), sets $\delta =0$,
thus establishing rotation additivity:

\begin{equation}
\left| \Psi \left( t\right) \right\rangle =\cos \left( \vartheta +\varphi
\right) \left| 0\right\rangle _{r}\left| 1\right\rangle _{s}+\sin \left(
\vartheta +\varphi \right) \left| 1\right\rangle _{r}\left| 0\right\rangle
_{s}.
\end{equation}

This is the behaviour required from a ``good'' transmission. The rotation of
qubit $r$ is identically transmitted to the other qubit $s$. In fact

\begin{equation}
Tr_{r}\left( \left| \Psi \left( t\right) \right\rangle \left\langle \Psi
\left( t\right) \right| \right) =\rho _{s}\left( t\right) =\sin ^{2}\left(
\vartheta +\varphi \right) \left| 0\right\rangle _{s}\left\langle 0\right|
_{s}+\cos ^{2}\left( \vartheta +\varphi \right) \left| 1\right\rangle
_{s}\left\langle 1\right| _{s},
\end{equation}
where $\left| \Psi \left( t\right) \right\rangle $ is given by eq. (7).
Eigenvalues 0 and 1 are interchanged: of course we are using a transmission
where one qubit is the NOT\ of the other. See also ref. [9,10,11].

Interestingly, by simultaneously rotating also the other extremity of the
transmission by the same amount $\left[ Q_{s}\left( \omega t\right) \rho
_{s}\left( 0\right) Q_{s}^{^{^{\dagger }}}\left( \omega t\right) \right] $,
the same result is obtained (eq. 7). Actually, this means adding eq. (8) as
a condition, but this is of course redundant with respect to condition (ii) $%
-$ it is derived from conditions (i) and (ii). Whereas, two different
rotations of the two transmission extremities give an impossible
mathematical system; this resembles a rigid classical transmission.

It should be noted that a rotation $\varphi $ of qubit (part) $r$ under $%
A_{rs}$ projection, is equivalent to applying the {\em unitary} operator $%
Q\left( \varphi \right) $ to the overall state $\left| \Psi \left( t\right)
\right\rangle $:

\begin{center}
$Q\left( \varphi \right) \equiv \left( 
\begin{array}{cccc}
\cos \varphi & \sin \varphi & 0 & 0 \\ 
-\sin \varphi & \cos \varphi & 0 & 0 \\ 
0 & 0 & \cos \varphi & -\sin \varphi \\ 
0 & 0 & \sin \varphi & \cos \varphi
\end{array}
\right) ,$ with

$\left| 0\right\rangle _{r}\left| 1\right\rangle _{s}\equiv \left( 
\begin{array}{c}
1 \\ 
0 \\ 
0 \\ 
1
\end{array}
\right) $ and $\left| 1\right\rangle _{r}\left| 0\right\rangle _{s}\equiv
\left( 
\begin{array}{c}
0 \\ 
1 \\ 
1 \\ 
0
\end{array}
\right) .$
\end{center}

$Q\left( \varphi \right) $ brings the overall state from $\left| \Psi \left(
0\right) \right\rangle $ (eq. 6) to $\left| \Psi \left( t\right)
\right\rangle $ (eq. 7) without ever violating (for all $\varphi $) $A_{rs}$%
. It {\em cannot} be represented as the product of two transformations
operating separately on the two qubits. As a matter of fact, it can be seen
that the result of operating on either qubit is conditioned by the state of
the other. There is a form of conditional logic implicit in this unitary
transformation, which operates on the whole state in an irreducible way.
Finding $Q\left( \varphi \right) $ amounts to solving a problem.

We have thus ascertained a peculiar fact. Our operation on a part, {\em blind%
} to its effect on the whole, performed together with continuous $A_{rs} $
projection, generates a {\em unitary} {\em transformation} which is, so to
speak, {\em wise} to the whole state, to how it should be transformed
without violating $A_{rs}$. Of course $A_{rs}$ ends up commuting with the
resulting overall unitary propagator (shaped by it). This is clearly an
application of the divide and conquer strategy, whose advantages will become
clear in the next section.

\section{Quantum computation networks}

Let us consider the reversible Boolean network of fig. 4, fully deployed in
space. This is different from sequential computation, where the Boolean
network appears in the space-time diagram of the computation process. Time
is now orthogonal to the network lay-out (for a computation of such networks
by means of projectors, see ref. 8).

\medskip

\medskip

\medskip \medskip \medskip

\begin{center}
Fig. 4
\end{center}

The network nodes $t$, $u$, $v$ and $r$ make the input and the output of a
controlled NOT; $r$ and $s$ belong to a quantum mechanical transmission as
defined above.

This c-NOT is not the usual time sequential gate, as its input and output
coexist. It is a quantum object of four qubits, and four eigenstates which
map the Boolean relation imposed by the gate. The Hilbert space of the gate
states is thus 
\[
H_{g}=span\left\{ \left| 0\right\rangle _{t}\left| 0\right\rangle _{u}\left|
0\right\rangle _{v}\left| 0\right\rangle _{r},\left| 0\right\rangle
_{t}\left| 1\right\rangle _{u}\left| 0\right\rangle _{v}\left|
1\right\rangle _{r},\left| 1\right\rangle _{t}\left| 0\right\rangle
_{u}\left| 1\right\rangle _{v}\left| 1\right\rangle _{r},\left|
1\right\rangle _{t}\left| 1\right\rangle _{u}\left| 1\right\rangle
_{v}\left| 0\right\rangle _{r}\right\} . 
\]
Model Hamiltonians of similar elementary gates have been developed in ref
[12]. Of course these gates differ from the gates used in the sequential
approach, where inputs and outputs are successive states of the same
register.$^{\left[ 13,14,15,16\right] }$

The usual way of stating a problem with such a space-deployed network, is by
constraining {\em part} of the input and {\em part} of the output (actually,
there is no preferred input-output direction) and asking whether this
constrained network admits a solution. Let $u=1$ and $s=1$ be such partial
constraints. $u=1$ ($s=1$) propagates a {\em conditional} logical
implication from left to right (right to left). The logical implication is
conditioned by the possible values of the unconstrained part of the input
(output). In order to have a solution, these two propagations must be
matched, namely must generate a univocal set of values on all the nodes of
the network $-$ of course comprising the unconstrained input/output nodes.
Finding whether the network admits at least one match (one solution) is an
NP-complete problem in general. Naturally, it is the well known SAT\
(satisfiability) problem on a reversible Boolean network.

Possible collisions (mismatch) between the two propagations will be both
overcome and reconciled by the transmission, as it will become clear.

Let us assume that the network has just one solution (which is the case
here: $t=1,$ $u=1,$ $r=0,$ $v=1,$ $s=1$). The procedure to find it is as
follows. The output constraint is removed while an arbitrary value, say $t=0$%
, is assigned to the unconstrained part of the input. The logical
propagation of this input toward the output yields $t=0$, $u=1$, $r=1$, $s=0$
($v=t$ will be disregarded). This computation is performed off line in
polynomial time. It serves to specify the initial state in which the network
must be prepared:

\[
\left| \Psi \left( 0\right) \right\rangle =\left| 0\right\rangle _{t}\left|
1\right\rangle _{u}\left| 1\right\rangle _{r}\left| 0\right\rangle _{s}; 
\]
of course this state satisfies $A_{rs}$. However, qubit $s$ is in the
eigenstate $\left| 0\right\rangle _{s}$ rather than $\left| 1\right\rangle
_{s}$ (the output constraint). It will be turned into $\left| 1\right\rangle
_{s}$ {\em under} $A_{rs}$ projection, by applying the continuous
transformation ${\bf 1}\rho _{u}\left( 0\right) {\bf 1}Q_{s}\left( \omega
t\right) \rho _{s}\left( 0\right) Q_{s}^{\dagger }\left( \omega t\right) $,
with $t$ going from 0 to $\frac{\pi }{2\omega }$. While $\rho _{s}$ is
rotated, $\rho _{u}=\left| 1\right\rangle _{u}\left\langle 1\right| _{u}$ is
kept fixed. This transformation operates on the network Hilbert space $H_{n}$
defined as $H_{n}=H_{g}\otimes H_{s}$, where $H_{s}=span\left\{ \left|
0\right\rangle _{s},\left| 1\right\rangle _{s}\right\} $. Note that all
states of $H_{n}$ already satisfy the gate, not necessarily the transmission.

At any time $t$, the state of the network is obtained by submitting a free
normalized state $\left| \Psi \left( t\right) \right\rangle $ of $H_{n}$ to
the conditions:

\begin{enumerate}
\item[(i)]  $A_{rs}\left| \Psi \left( t\right) \right\rangle =\left| \Psi
\left( t\right) \right\rangle $

\item[(ii)]  $Tr_{t,u,r}\left( \left| \Psi \left( t\right) \right\rangle
\left\langle \Psi \left( t\right) \right| \right) =\rho _{s}\left( t\right)
=Q_{s}\left( \omega t\right) \rho _{s}\left( 0\right) Q_{s}^{\dagger }\left(
\omega t\right) $

\item[(iii)]  $Tr_{t,r,s}\left( \left| \Psi \left( t\right) \right\rangle
\left\langle \Psi \left( t\right) \right| \right) =\rho _{u}\left( 0\right)
=\left| 1\right\rangle _{u}\left\langle 1\right| _{u}$

\item[(iv)]  the distance between the vectors before and after projection
should be minimum.
\end{enumerate}

This yields at time $t$:

\[
\left| \Psi \left( t\right) \right\rangle =\cos \varphi \left|
0\right\rangle _{t}\left| 1\right\rangle _{u}\left| 1\right\rangle
_{r}\left| 0\right\rangle _{s}+e^{i\delta }\sin \varphi \left|
1\right\rangle _{t}\left| 1\right\rangle _{u}\left| 0\right\rangle
_{r}\left| 1\right\rangle _{s}, 
\]
with $\varphi =\omega t$, which is readily checked. Condition (iv) and the
network initial state set $\delta =0$. Thus 
\begin{equation}
\left| \Psi \left( t\right) \right\rangle =\cos \varphi \left|
0\right\rangle _{t}\left| 1\right\rangle _{u}\left| 1\right\rangle
_{r}\left| 0\right\rangle _{s}+\sin \varphi \left| 1\right\rangle _{t}\left|
1\right\rangle _{u}\left| 0\right\rangle _{r}\left| 1\right\rangle _{s}
\end{equation}

For $\varphi =\frac{\pi }{2}$, one obtains

\[
\left| \Psi \left( \frac{\pi }{2\omega }\right) \right\rangle =\left|
1\right\rangle _{t}\left| 1\right\rangle _{u}\left| 0\right\rangle
_{r}\left| 1\right\rangle _{s}, 
\]
namely the solution.

The ``wise'' unitary transformation (9) (to how to behave on the whole)
brings the state of the network from satisfying only the input to satisfying
both the input and the output constraints. It is obtained by ``blindly''
operating on {\em divided parts} of the network, but under $A_{rs}$
projection. The latter is the {\em conquering} agent.

It is worth analysing the character of this computation process. We will
consider a generic Boolean network where each wire comprises (without loss
of generality) a NOT\ function, and is implemented by a transmission.

First, let us assume that the network has exactly one solution. The network
eigenstates $-$ tensor products of eigenstates of the network parts $-$
inherently satisfy all gates (by definition, a gate is a part whose
eigenstates map the gate logical input-output function). Network preparation
is described by just one tensor product (as in the example of fig. 4).
Beyond the gates, this satisfies input constraints and all transmissions,
not the output constraint (but for a lucky chance).

Then, we start operating on the density matrices of those input/output
qubits which must satisfy the external constraints, {\em under} {\em the} $%
A_{rs}$ {\em projections} (one for each transmission), in order to keep the
constrained inputs unaltered, while bringing the output in match with its
constraint. This computation process works on the basis vectors of the
network Hilbert space, which satisfy all gates, and cannot originate a
tensor product which violates an $A_{rs}$ symmetry. In counterfactual
reasoning, if there had been such a ``bad'' tensor product, its amplitude
would have been canceled by continuous $A_{rs}$ projection and, through
renormalization, would have gone to feed the amplitudes of the ``good''
tensor products. Computation becomes thus a {\em unitary} {\em evolution} of
the overall network state, {\em driven} by the operations on some parts (on
the input/output qubits which must satisfy the external constraints) and 
{\em shaped} by $A_{rs}$ projections.

As a result of this process, $A_{rs}$ symmetries become constants of motion
which commute with the propagation at all times. They are also pairwise
commuting, being applied to disjoint Hilbert spaces. However, the cause
should not be confused with the effect. $A_{rs}$ projections shape or forge
the unitary evolution with which they commute. This, according to the
projection interpretation of particle statistics (Section III).

It is clear from this picture that the two propagations of conditional
logical implication, from input to output and vice-versa, never originate a
collision (a tensor product where a transmission is in an inconsistent
state). The amplitudes of collisions must be zero because of continuous
projection. Computation can never go into a deadlock. It is of course a {\em %
unitary evolution} toward the solution, driven, as we have seen, by
operating on the parts and shaped by projecting the whole.

If the network admits no solution, conditions (i) through (iv) make up an
impossible system. Measuring the network final state $-$ when the operations
on the parts are completed $\left( t=\frac{\pi }{2\omega }\right) $ $-$
gives some random result which is not a solution. This is checkable in
polynomial time and tells that the network is not satisfiable.

If the network admits many solutions, the final state can be a linear
combination thereof. Which linear combination, depends on the network
initial state through condition (iv). This can be seen through numerical
calculation. However this is irrelevant since measurement anyhow gives one
solution (that it is a solution is checkable in polynomial time).

We should better clarify how parts should be connected by transmissions.
Fig. 5 illustrates the appropriate arrangement (the network of fig. 4 was
simplified). All elementary gates and the constrained input/output qubits
are interconnected through transmissions. Namely any network {\em wire}
connecting nodes say $r$ and $s$, becomes a {\em transmission}, or projector 
$A_{rs}$.

It should be made clear that this is not necessary, but useful. Two or more
elementary gates could be interconnected without the interposition of
transmissions. However, this would require ``building'' an exponentially
more complex gate and the number of its eigenstates would be the product of
the number of eigenstates of the individual gates. Transmissions are exactly
meant to reduce network complexity (exponentially with the number of
transmissions). Clearly, they introduce a divide and conquer strategy.This
quantum form of reductionism would be more rewarding than (the known)
classical reductionism. At this abstract level, it would make NP-complete $%
\equiv $ P. Of course we are applying reverse engineering and we have just
highlighted a nice-to-have feature, namely the $A_{rs}$ projection. This
raises the problem whether this feature is physical. Until now, we have seen
that these projections share the nature of particle statistics symmetries,
under a special interpretation thereof.

\section{Induced symmetry}

We will show that the ``artificial'' symmetry $A_{rs}$ of a quantum
mechanical transmission is an epiphenomenon of fermionic antisymmetry in a
special physical situation. This is generated by submitting a couple of
identical fermions 1 and 2 to a suitable Hamiltonian $^{[9]}$. We assume
that each fermion has two compatible, binary degrees of freedom $\chi $ and $%
\lambda $. Just for the sake of visualization, we can assume that each
fermion is a spin $1/2$ particle and can occupy one of either two sites of a
spatial lattice. In this case $\chi $ becomes the particle spin component $%
\sigma _{z}=0$ (down), $1$ (up) and $\lambda =r,s$ the label of the site
occupied by the particle.

Therefore the Hilbert space of the spatial states of the two fermions is:

\[
H_{\lambda }=span\left\{ \left| r\right\rangle _{1}\left| r\right\rangle
_{2},\left| s\right\rangle _{1}\left| s\right\rangle _{2},\frac{1}{\sqrt{2}}%
\left( \left| r\right\rangle _{1}\left| s\right\rangle _{2}-\left|
s\right\rangle _{1}\left| r\right\rangle _{2}\right) ,\frac{1}{\sqrt{2}}
\left( \left| r\right\rangle _{1}\left| s\right\rangle _{2}+\left|
s\right\rangle _{1}\left| r\right\rangle _{2}\right) \right\} , 
\]
where $\left| r\right\rangle _{1}\left| r\right\rangle _{2}$ means that
particle 1 is in site $r$, particle 2 is also in site $r$, etc. The Hilbert
space of the spin states of the two fermions is:

\[
H_{\chi }=span\left\{ \left| 0\right\rangle _{1}\left| 0\right\rangle
_{2},\left| 1\right\rangle _{1}\left| 1\right\rangle _{2},\frac{1}{\sqrt{2}}%
\left( \left| 0\right\rangle _{1}\left| 1\right\rangle _{2}+\left|
1\right\rangle _{1}\left| 0\right\rangle _{2}\right) ,\frac{1}{\sqrt{2}}
\left( \left| 0\right\rangle _{1}\left| 1\right\rangle _{2}-\left|
1\right\rangle _{1}\left| 0\right\rangle _{2}\right) \right\} , 
\]
$\left| 0\right\rangle _{1}\left| 0\right\rangle _{2}$ means that the $%
\sigma _{z}$ component of particle 1 is 0 (down), etc. The overall Hilbert
space $H_{\lambda \chi }=H_{\lambda }\otimes H_{\chi }$ has 16 basis
vectors. The following is the list of the states which do not violate
statistics. States are represented in first and second quantization and,
when applicable, in qubit notation:

\begin{itemize}
\item  Both particles in the same site, spin (necessarily) in the singlet
state:
\end{itemize}

$\left| a\right\rangle =\frac{1}{\sqrt{2}}\left( \left| 0\right\rangle
_{1}\left| 1\right\rangle _{2}-\left| 1\right\rangle _{1}\left|
0\right\rangle _{2}\right) \left| r\right\rangle _{1}\left| r\right\rangle
_{2}=a_{0r}^{\dagger }\ a_{1r}^{\dagger }\left| 0\right\rangle ,$

$\left| b\right\rangle =\frac{1}{\sqrt{2}}\left( \left| 0\right\rangle
_{1}\left| 1\right\rangle _{2}-\left| 1\right\rangle _{1}\left|
0\right\rangle _{2}\right) \left| s\right\rangle _{1}\left| s\right\rangle
_{2}=a_{0s}^{\dagger }\ a_{1s}^{\dagger }\left| 0\right\rangle .$

\begin{itemize}
\item  One particle per site, antisymmetrical spatial state vector, spin
(necessarily) in a triplet state:
\end{itemize}

$\left| c\right\rangle =\frac{1}{\sqrt{2}}\left| 0\right\rangle _{1}\left|
0\right\rangle _{2}\left( \left| r\right\rangle _{1}\left| s\right\rangle
_{2}-\left| s\right\rangle _{1}\left| r\right\rangle _{2}\right)
=a_{0r}^{\dagger }\ a_{0s}^{\dagger }\left| 0\right\rangle =\left|
0\right\rangle _{r}\left| 0\right\rangle _{s},$

$\left| d\right\rangle =\frac{1}{\sqrt{2}}\left| 1\right\rangle _{1}\left|
1\right\rangle _{2}\left( \left| r\right\rangle _{1}\left| s\right\rangle
_{2}-\left| s\right\rangle _{1}\left| r\right\rangle _{2}\right)
=a_{1r}^{\dagger }\ a_{1s}^{\dagger }\left| 0\right\rangle =\left|
1\right\rangle _{r}\left| 1\right\rangle _{s},$

$\left| e\right\rangle =\frac{1}{2}\left( \left| 0\right\rangle _{1}\left|
1\right\rangle _{2}+\left| 1\right\rangle _{1}\left| 0\right\rangle
_{2}\right) \left( \left| r\right\rangle _{1}\left| s\right\rangle
_{2}-\left| s\right\rangle _{1}\left| r\right\rangle _{2}\right) =\frac{1}{%
\sqrt{2}}\left( a_{0r}^{\dagger }\ a_{1s}^{\dagger }+a_{1r}^{\dagger }\
a_{0s}^{\dagger }\right) \left| 0\right\rangle =$

$\frac{1}{\sqrt{2}}\left( \left| 0\right\rangle _{r}\left| 1\right\rangle
_{s}+\left| 1\right\rangle _{r}\left| 0\right\rangle _{s}\right) .$

\begin{itemize}
\item  One particle per site, symmetrical spatial state vector, spin
(necessarily) in the singlet state:
\end{itemize}

$\left| f\right\rangle =\frac{1}{2}\left( \left| 0\right\rangle _{1}\left|
1\right\rangle _{2}-\left| 1\right\rangle _{1}\left| 0\right\rangle
_{2}\right) \left( \left| r\right\rangle _{1}\left| s\right\rangle
_{2}+\left| s\right\rangle _{1}\left| r\right\rangle _{2}\right) =\frac{1}{%
\sqrt{2}}\left( a_{0r}^{\dagger }\ a_{1s}^{\dagger }-a_{1r}^{\dagger }\
a_{0s}^{\dagger }\right) \left| 0\right\rangle =$

$\frac{1}{\sqrt{2}}\left( \left| 0\right\rangle _{r}\left| 1\right\rangle
_{s}-\left| 1\right\rangle _{r}\left| 0\right\rangle _{s}\right) .$

\medskip

\noindent Creation/annihilation operators form the algebra:

\begin{equation}
\left\{ a_{i}^{\dagger },a_{j}^{\dagger }\right\} =\left\{
a_{i},a_{j}\right\} =0,\ \ \left\{ a_{i}^{\dagger },a_{j}\right\} =\delta
_{i,j}.
\end{equation}

Now we introduce the Hamiltonian

\[
H_{rs}=E_{a}\left| a\right\rangle \left\langle a\right| +E_{b}\left|
b\right\rangle \left\langle b\right| +E_{c}\left| c\right\rangle
\left\langle c\right| +E_{d}\left| d\right\rangle \left\langle d\right| , 
\]
or, in second quantization,

\[
H_{rs}=-\left( E_{a}\ a_{0r}^{\dagger }\ a_{1r}^{\dagger
}a_{0r}a_{1r}+E_{b}\ a_{0s}^{\dagger }\ a_{1s}^{\dagger }a_{0s}a_{1s}+E_{c}\
a_{0r}^{\dagger }\ a_{0s}^{\dagger }a_{0r}a_{0s}+E_{d}\ a_{1r}^{\dagger }\
a_{1s}^{\dagger }a_{1r}a_{1s}\right) , 
\]

with $E_{a}$, $E_{b}>E_{c}$, $E_{d}\geq E$ discretely above 0.

This leaves us with two degenerate ground eigenstates:

\[
\left| e\right\rangle =\frac{1}{\sqrt{2}}\left( \left| 0\right\rangle
_{r}\left| 1\right\rangle _{s}+\left| 1\right\rangle _{r}\left|
0\right\rangle _{s}\right) \text{ and }\left| f\right\rangle =\frac{1}{\sqrt{%
2}}\left( \left| 0\right\rangle _{r}\left| 1\right\rangle _{s}-\left|
1\right\rangle _{r}\left| 0\right\rangle _{s}\right) . 
\]
Alternatively, their linear combinations $\left| 0\right\rangle _{r}\left|
1\right\rangle _{s}$ and $\left| 1\right\rangle _{r}\left| 0\right\rangle
_{s}$ can be used as the two orthogonal ground eigenstates.

The generic ground state is thus:

\begin{equation}
\left| \Psi \right\rangle =\alpha \left| 0\right\rangle _{r}\left|
1\right\rangle _{s}+\beta \left| 1\right\rangle _{r}\left| 0\right\rangle
_{s}\text{, with }\left| \alpha \right| ^{2}+\left| \beta \right| ^{2}=1.
\end{equation}
It can be seen that $\left| \Psi \right\rangle $ satisfies $A_{rs}$ symmetry.

Let $A_{12}=1-P_{12}$ and renormalization be the antisymmetrization
projector, where $P_{12}$ permutes particles 1, 2. Then $A_{12}\left| \Psi
\right\rangle =\left| \Psi \right\rangle $ means that $\left| \Psi
\right\rangle $ is antisymmetric.

Due to the anticommutation relations (10), $A_{12}\left| 0\right\rangle
_{r}\left| 1\right\rangle _{s}=\left| 0\right\rangle _{r}\left|
1\right\rangle _{s}$ and $A_{12}\left| 1\right\rangle _{r}\left|
0\right\rangle _{s}=\left| 1\right\rangle _{r}\left| 0\right\rangle _{s}$.
Also, $A_{12}\left| 0\right\rangle _{r}\left| 0\right\rangle _{s}=\left|
0\right\rangle _{r}\left| 0\right\rangle _{s}$ and $A_{12}\left|
1\right\rangle _{r}\left| 1\right\rangle _{s}=\left| 1\right\rangle
_{r}\left| 1\right\rangle _{s}$, without forgetting that these are excited
states.

We will show that a quantum mechanical transmission can be implemented by
suitably operating on the ground state (11). Without significant
restriction, we assume that the initial (``symmetrical'') state of the
transmission is given by eq. (6), repeated here for convenience: 
\[
\left| \Psi \left( 0\right) \right\rangle =\cos \vartheta \left|
0\right\rangle _{r}\left| 1\right\rangle _{s}+\sin \vartheta \left|
1\right\rangle _{r}\left| 0\right\rangle _{s},
\]
Then we apply transformation (5) to qubit $r$: 
\[
\rho _{r}\left( t\right) =Q_{r}\left( \omega t\right) \rho _{r}\left(
0\right) Q_{r}^{\dagger }\left( \omega t\right) .
\]

Let $\left| \Psi \left( t\right) \right\rangle $ be a free normalized vector
of $H_{\lambda \chi }$. The transmission state at time $t$ is obtained by
submitting $\left| \Psi \left( t\right) \right\rangle $ to the following
mathematically simultaneous conditions:

\begin{enumerate}
\item[i)]  $A_{12}\left| \Psi \left( t\right) \right\rangle =\left| \Psi
\left( t\right) \right\rangle ,$

\item[ii)]  $\rho _{r}\left( t\right) =Tr_{s}\left( \left| \Psi \left(
t\right) \right\rangle \left\langle \Psi \left( t\right) \right| \right)
=\cos ^{2}\left( \vartheta +\varphi \right) \left| 0\right\rangle
_{r}\left\langle 0\right| _{r}+\sin ^{2}\left( \vartheta +\varphi \right)
\left| 1\right\rangle _{r}\left\langle 1\right| _{r},$

\item[iii)]  the distance between the vectors before and after reduction
must be minimum

\item[iv)]  the expected energy of the transmission in $\left| \Psi \left(
t\right) \right\rangle :\left\langle \xi \left( t\right) \right\rangle
=\left\langle \Psi \left( t\right) \right| H_{rs}\left| \Psi \left( t\right)
\right\rangle ,$ must be minimum
\end{enumerate}

\noindent It is readily seen that the solution of this system is still $%
\left| \Psi \left( t\right) \right\rangle $ of eq. (7): 
\[
\left| \Psi \left( t\right) \right\rangle =\cos \left( \vartheta +\varphi
\right) \left| 0\right\rangle _{r}\left| 1\right\rangle _{s}+\sin \left(
\vartheta +\varphi \right) \left| 1\right\rangle _{r}\left| 0\right\rangle
_{s}.
\]
Simultaneous satisfaction of (i), i.e. fermionic antisymmetry viewed as a
projection, and (iv) (which is satisfied by $\left\langle \xi \left(
t\right) \right\rangle =0$) originates the projection $A_{rs}\left| \Psi
\left( t\right) \right\rangle =\left| \Psi \left( t\right) \right\rangle $,
as is readily seen. Therefore, if $\left\langle \xi \left( t\right)
\right\rangle $ remains zero, namely the operation on qubit $r$ is performed 
{\em adiabatically}, we obtain a quantum mechanical transmission as
described in section IV. This transmission undergoes a unitary
transformation driven by $Q_{r}\left( \omega t\right) $ and shaped by $A_{rs}
$ (i.e. $A_{12}$ {\em and} $\left\langle \xi \left( t\right) \right\rangle =0
$).

The operation on qubit $r$ should be gentle enough to be adiabatic. It
should be noted that this is a local problem, whereas network size is
irrelevant. The consequences of the operation on $r$ are propagated
throughout the network by means of the transmissions, therefore by way of
interference (destructive due to projection and constructive due to
renormalization). We should note that the tensor products $\left|
0\right\rangle _{r}\left| 0\right\rangle _{s}$ and $\left| 1\right\rangle
_{r}\left| 1\right\rangle _{s}$ that would be projected off since they
violate particle statistics (Section V), are {\em not} the excited states $%
\left| c\right\rangle $ and $\left| d\right\rangle $, which satisfy $A_{12}$%
. They would be instead the symmetrical states $\left| 0\right\rangle
_{r}\left| 0\right\rangle _{s}=\frac{1}{\sqrt{2}}\left| 0\right\rangle
_{1}\left| 0\right\rangle _{2}\left( \left| r\right\rangle _{1}\left|
s\right\rangle _{2}+\left| s\right\rangle _{1}\left| r\right\rangle
_{2}\right) $ and $\left| 1\right\rangle _{r}\left| 1\right\rangle _{s}=%
\frac{1}{\sqrt{2}}\left| 1\right\rangle _{1}\left| 1\right\rangle _{2}\left(
\left| r\right\rangle _{1}\left| s\right\rangle _{2}+\left| s\right\rangle
_{1}\left| r\right\rangle _{2}\right) $, which do not satisfy $A_{12}$. As a
matter of fact, such states would be, so to speak, immediately projected off
by statistics. The two kinds of states (antisymmetrical and symmetrical)
have the same qubit notations and density matrices.

Given that the symmetric states cannot exist, this is counterfactual
reasoning, which might help understanding; the important thing remains that
conditions (i) through (iv) yield the solution (7). To ensure that the
operation on qubit $r$ is performed adiabatically, the transmission density
matrix should be kept on ground.

However, let us examine the case that the operation in question is not fully
adiabatic. An antisymmetric, excited state of the form $\gamma \left|
0\right\rangle _{r}\left| 0\right\rangle _{s}+\delta \left| 1\right\rangle
_{r}\left| 1\right\rangle _{s}$ would appear in {\em superposition} with the
transmission ground state: 
\[
\left| \Psi \left( \frac{\pi }{2\omega }\right) \right\rangle =\alpha \left|
0\right\rangle _{r}\left| 1\right\rangle _{s}+\beta \left| 1\right\rangle
_{r}\left| 0\right\rangle _{s}+\gamma \left| 0\right\rangle _{r}\left|
0\right\rangle _{s}+\delta \left| 1\right\rangle _{r}\left| 1\right\rangle
_{s}, 
\]
(the antisymmetrical, excited states where both particles are in the same
site can be dealt with in a similar way). Let us assume this to be the state
at the end of the operations. The probability that the result of measurement
gives a transmission ``malfunction'' ($\left| 0\right\rangle _{r}\left|
0\right\rangle _{s}$ or $\left| 1\right\rangle _{r}\left| 1\right\rangle
_{s} $) is $q=\left| \gamma \right| ^{2}+\left| \delta \right| ^{2}.$ The
expected transmission energy is $\left\langle \xi \left( t\right)
\right\rangle =\left| \gamma \right| ^{2}E_{c}+\left| \delta \right|
^{2}E_{d}$. In order to keep $q$ small, it should be $\left\langle \xi
\left( t\right) \right\rangle \ll E_{c},E_{d}.$ Since this computation is 
{\em reversible, }namely it does not dissipate free energy (the result of 
{\em driving and shaping} is a unitary transformation), it seems that $%
\left\langle \xi \left( t\right) \right\rangle $ can be kept as close to
zero as desired.

However, in order to keep network complexity down, the number of
transmissions should grow linearly with network size. If the probability of
a transmission ``malfunction'' remains constant, the probability that there
are no malfunctions at network level (in the measurement outcome) would
exponentially decrease with network size $-$ in this case the SAT\ problem
would remain NP-complete.

But transmissions do relax toward ground state, moreover independently of
each other, since they also decouple the parts of the network (then
re-coupled by projections). The pace of computation $\left( \omega \right) $
could be slowed down so that computation is continuously caught up by the
relaxation process of the transmissions. At any time, there should be a
fixed desired probability $p_{N}$ that a measurement would find all
transmissions in the ground state.

Let us assume that $p=1-e^{-\sigma t}$, with $\sigma >0$, is the probability
of finding an individual transmission in the ground state at time $t$ (this
exponential law holds when relaxation has reached a constant rate). Under
this assumption, it can be shown analytically that, for a given $p_{N}$,
computation time grows polynomially with network size. This would mean
NP-complete $\equiv P.$

However, for the time being these are conjectural discussions: an
implementation model would be needed in order to move to a less speculative
analysis.

Let us now address the problem of creating many transmissions, namely an $%
H_{rs}$ Hamiltonian per network wire $r,s$ (fig. 5). These Hamiltonians
operate on disjoint pairs of qubits. Viewed as projectors (when $%
\left\langle \xi \left( t\right) \right\rangle =0$), they are pairwise
commuting in spite of the fact that qubits belonging to different
transmissions are mutually bounded by the network gates. Actually, all $%
A_{rs}$ projectors commute with the propagator of the network state (shaped
by the $A_{rs}$).

\section{Discussion}

The notion of applying a particle statistics symmetry (seen as a projector)
to divide the quantum whole into parts without clipping its richness,
introduces an engineering (reductionistic) perspective in the design of
quantum mechanisms. For the time being, the development of this idea remains
at an abstract level. Finding model Hamiltonians which implement the
Hermitean matrix of Section VI could possibly be the next step.

The form of computation propounded seems to imply a somewhat deeper
interpretation of the notion of time-reversibility. Indeed the very notion
of quantum computation was born as an evolution of the notion of reversible
computation$^{\left[ 17,18,19,20,21\right] }$. Now we further conjecture
that any quantum computation speed-up is related to the notion of the
coexistence of forward and backward in time computation (or propagation).

This can be shown both in the traditional approach of sequential quantum
computation (where the Boolean network appears in the time-diagram of the
computation process) and in the approach propounder here. 

Let us consider sequential computation first. Here computation speed-up can
be related to the notion of performing inverse computation (of a hard to
invert function) by using direct computation in a time reversed way (which
of course requires computation reversibility). In this way, the time
interval used by inverse computation is the same of direct computation.

For example, with Simon's algorithm, given a (hard to invert) function $%
f:B^{n}\rightarrow B^{n}$, 2-to-1 with periodicity $r$, $r$ can always be
found in polynomial time. An essential point of this algorithm is measuring $%
f\left( x\right) $ in the entangled state.

\[
\sum_{x}\left| x\right\rangle \left| f\left( x\right) \right\rangle . 
\]
Say this yields $f\left( x\right) =k.$ This gives, in the left register

\begin{equation}
\frac{1}{\sqrt{2}}\left( \left| f^{-1}\left( k\right) \right\rangle +\left|
f^{-1}\left( k\right) +r\right\rangle \right) .
\end{equation}

Then, after applying the Hadamard transform, measurement of the left
register and some repetitions of the entire process bring $r$ out in the
episystem. However, it can be argued that the quantum speed-up is already in
the intermediate result (12). Roughly speaking, if this were the printout of
a classical computer (with $f^{-1}\left( k\right) $ and $r$ substituted by
the appropriate numerical assignments), computation time would be
exponential.

Let us consider fig. 6. Measurement of the right register (right-top of
figure 6), {\em after} computation, gives no information since the result is
random. This result, $f\left( x\right) =k$, goes backward in time along the
right branch $r$ to the interaction region (the cloud), where the {\em %
direct function} is computed. Since the result of measurement now goes back
in time, the {\em inverse function} is computed (which would require
exponential time in classical computation). Naturally, the time interval
required is that for computing the direct function. This effectively implies
the coexistence of forward and backward in time computation. The result of
inverse computation is state (12) in the left register, which goes forward
in time along the left branch. Further processing brings the symbol $r$ out
in the episystem. This algorithm clearly computes the inverse function
through a time inversion of direct computation.

By the way, this is no time-travel, no information can go to the past. The
result of measurement is random and provides no information. Thus, there is
no information in the final state of computation. Furthermore, reversible
computation$^{\left[ 17,18,19\right] }$ neither destroys nor creates
information. To sum up, there is no information whatsoever that can go back
in time.

This is much similar to EPR\ correlations. EPR entanglement can be seen as
an elementary computation. One can still use fig. 6 with $\sum_{x}\left|
x\right\rangle \otimes \left| f\left( x\right) \right\rangle =\left|
0\right\rangle \left| 1\right\rangle -\left| 1\right\rangle \left|
0\right\rangle $ (normalization is disregarded). Measurement on, say, the
right particle originates a random result, this is first propagated backward
in time to the region of the interaction, then forward in time to the left
particle, which carries that result to the left measurement. This EPR
interpretation has been formalized by using a two-way propagation model in
[22].

This interpretation of the quantum computation speed-up can be extended to
all quantum algorithms found so far$^{\left[ 23\right] }$ $-$ readily by
using the unified model developed by A. Ekert$^{\left[ 24\right] }$.

Two-way computation/propagation is also an interpretation of the computation
model propounded in this work. The {\em state} of the transmission $r,s$ at
some time $t$ must change due to the operations performed on a {\em part of
it}. The new overall state is forged by $A_{rs}$ projection. This projection
gives two possible outcomes: one in $H_{s}$ (satisfying $A_{rs}$), the other
in the orthogonal subspace $H_{s\perp }$ (violating $A_{rs}$). Because of
continuous projection, the probability of violating $A_{rs}$ is rounded off
to sharp zero.

This rounding off is implied in the mathematically simultaneous application
of (i) operation on a part and (ii) projection of the whole on a {\em %
predetermined} {\em outcome} (conditions (i) through (iv) of Section VI). In
other words, the choice between ending up in $H_{s}$ or $H_{s\perp }$ is
predetermined by the requirement that the {\em projection} outcome will
satisfy a certain final condition ($A_{rs}$ symmetry). Projection on a
predetermined outcome ($H_{s}$) would be ``magic'' if the state to be
projected had cumulated a discrete distance from $H_{s}$. A condition set in
the future would determine a choice in the past. It is less ``magic''
because projection is continuous. Still, computation speed-up is essentially
related to it. So to speak, this is equivalent to always guessing the right
choice in a decision process $-$ in pruning a decision tree. This overcomes
the ``blindness'' of the operations on the parts and makes the evolution
wise to how to reach (unitarily) the problem solution. This also implies a
propagation which is elusively (without time-travel) driven by {\em both}
the initial and the final conditions$^{\left[ 22,25\right] }$. See also ref. 
$\left[ 9\right] .$ Other literature on two-way propagation models can be
found in refs. [26,27,28].

This research has been supported by Elsag Bailey a Finmeccanica company.
Thanks are due to A. Ekert, D. Finkelstein, L. Levitin, C. Macchiavello and
T. Toffoli for useful suggestions.

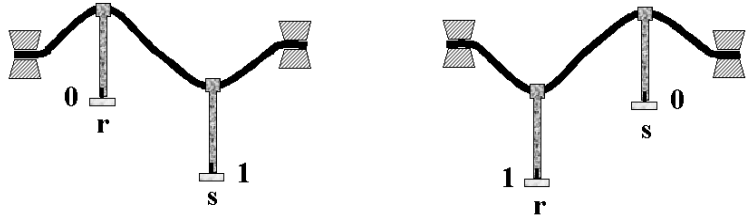
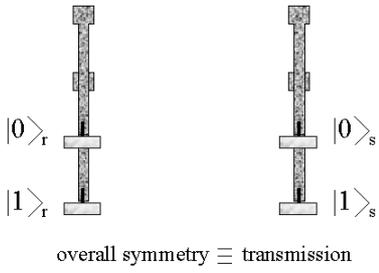
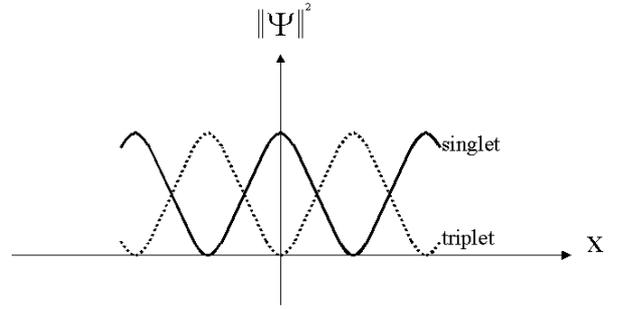
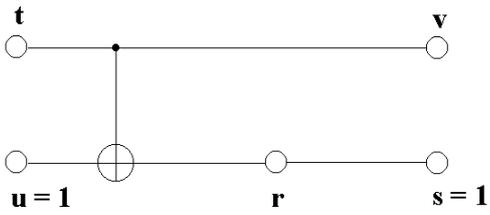
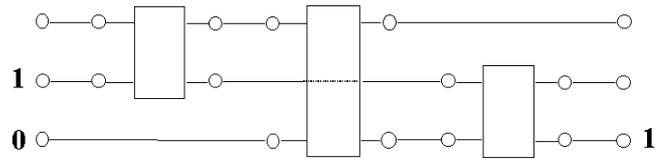
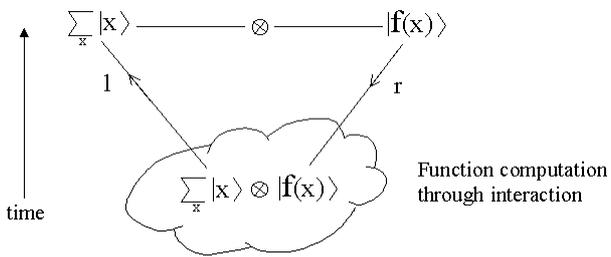
\end{document}